\documentclass[twocolumn,english,aps,graphicx,amsmath,showpacs]{revtex4}
\usepackage[T1]{fontenc}
\usepackage[latin1]{inputenc}
\usepackage{babel}
\usepackage{graphics}

\begin{document}

\title{A giant controllable gap in the optical spectrum of a semiconductor laser subject to intense feedback with rotated polarization.}
\author{Noam Gross, Zav Shotan, Tal Galfsky, and Lev Khaykovich}
\affiliation{Department of Physics, Bar-Ilan University, Ramat-Gan,
52900 Israel,}

\begin{abstract}
A semiconductor laser subject to delayed optical feedback is
investigated in the limit of extremely intense feedback power. In a
range of feedback polarization rotation angles the emission spectra
of the laser reveal a giant gap with width of more than a terahertz.
The position of the gap and its width are shown to be regulated by
means of feedback polarization rotation angle. We demonstrate that a
theoretical approach, based on carrier density grating induced
potential, explains our experimental results.
\end{abstract}

\maketitle

A semiconductor laser (SL) subject to optical feedback is a most
popular playground for fundamental understanding of physics of delay
systems. A rich variety of nonlinear dynamics phenomena has been
observed in different experimental setups and new dynamical states
have been recently reported \cite{Hohl99,Heil01,Ushakov04}. Despite
the complex behavior generally observed in delayed systems, a number
of schemes were proposed and experimentally realized to govern the
induced dynamics. Optical feedback from an external cavity was shown
to control unstable steady states in a multisection single-mode
laser \cite{Schikora06}. Another notable all optical control
mechanism was demonstrated, by means of a frequency dependent
feedback, to produce controlled frequency oscillations
\cite{Vemuri04} and frequency bistability \cite{Farias05}. In most
of the above cases a single-mode Lang-Kobayshi rate equation
approach \cite{LangKobayshi} was sufficient to describe the
experimental results.

Multi-mode dynamics of a SL subject to optical feedback is generally
less regulative and its theoretical analysis is more demanding.
Different aspects of this complicated problem have been investigated
experimentally \cite{Uchida01,Peil06} and several theoretical models
were suggested to address these results
\cite{Viktorov00,Masoller05,Serrat03}. Exceptional features
characterize the multi-mode dynamics, such as antiphase intensity
oscillations of the longitudinal modes (in a regime of weak
feedback) \cite{Uchida01,Masoller05} and the appearance of unusually
broadband optical spectrum \cite{Peil06}.

In this Letter we show a remarkably controllable and unexplored
regime of nonlinear multi-mode dynamics. We investigate a high power
SL subject to an optical feedback in a long external cavity
configuration. Our setup is unique in that it employs an extremely
intense and variably rotated polarization feedback. In fact, the
feedback is so strong that the external cavity dominates the
internal one which is in clear violation of the Lang-Kobayshi rate
equations assumption. Under these conditions and in a specific range
of polarization rotation angles, the laser emission is characterized
by an unusual broadband optical spectrum (up to $\sim3$ nm) and thus
by a short coherence length ($\sim100$ $\mu$m). Optical spectra
measurements reveal a distant two peak structure roughly centered at
the wavelength of the solitary laser. The splitting in the spectrum
is more than one terahertz wide, which exceeds by two orders of
magnitude all previously reported optical spectrum modulations for
similar single-feedback conditions \cite{Ushakov04,Vemuri04,Li98}.
We show the ability to control the width and position of the
splitting by means of feedback polarization angle rotation. In
closely related set of coherence length measurements we identify
multi-peak structures which correspond to a long wavelength
modulation of the gain medium spectrum.

To model the experimental results we use a theoretical approach that
assumes slow free carrier diffusion rates and, thus, predicts the
formation of carrier density grating inside the laser cavity with
periodicity imposed by the standing wave of lasing modes
\cite{Carr01,Masoller05}. The grating induces periodic modulation of
the gain medium refractive index which opens a gap in the optical
spectrum of the laser. We calculate the band-gap structure of the
induced potential and find that its qualitative behavior and key
quantitative values explain our experimental results. Our ability to
change the gap position and its width indicates high level of
control over the strength of periodic potential via a rotation of
the feedback polarization angle.


\begin{figure}

{\centering \resizebox*{0.45\textwidth}{0.27\textheight}
{{\includegraphics{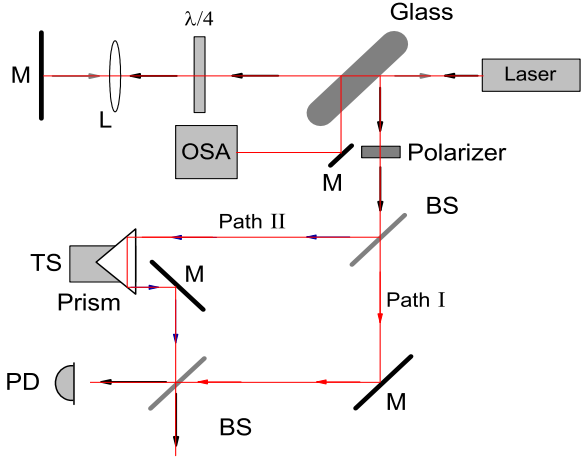}}}
\par}
\caption{\label{setup}A schematic representation of the experimental
setup. A $\lambda/4$ waveplate in the optical feedback path allows
for flexible polarization rotation to any angle. $4\%$ of light is
reflected to a Mach-Zender interferometer for coherence
length measurements. Another $4\%$ is reflected to an Optical
Spectrum Analyzer (OSA). BS - Beam Splitter; M - mirror; TS -
translation stage; PD - Photodetector.}
\end{figure}

In our experiments we use a Fabry-P\'{e}rot SL emitting 45 mW of
power at 663 nm (ML101J8). We operate the laser at maximum power
which requires the driving current to be well above the threshold
current. There is no special anti-reflection coating added to the
laser. The cavity length is $\sim$1 mm and the longitudinal mode
spacing is $\sim$45 GHz (refractive index of the gain medium is
$n=3.3$). An optical feedback is achieved by introducing a mirror
2.3 m away which simply reflects back the output power of the laser
(see Fig.\ref{setup}). A lens is located in front of the mirror to
improve modematching of the feedback beam profile onto the output
facet of the laser, therefore enhancing feedback power. We work at
maximum optical feedback level which corresponds to a reduction in
laser threshold current by 20$\%$ (measured in the regular feedback
configuration, i.e. with no polarization rotation). At this level
the external cavity dominates the internal one which is indicated in
dramatic reduction of feedback induced noise in laser output power
and narrowing of the optical spectrum. The optical feedback path
includes a quarter-wave plate to allow flexible polarization
rotation by any angle between 0$^{0}$ (same polarization) and
90$^{0}$ (orthogonal polarization). About $4\%$ of the output power,
split by a glass, is used as an input signal to an optical spectrum
analyzer. Another $4\%$ of the output power is used in a Mach-Zender
interferometer to measure coherence length of the laser, while
polarization of the light is cleaned at the entrance by a polarizer.
In one beam path of the interferometer a prism is installed on a
high precision piezo-motor translation stage which allows 20 mm scan
of the optical path difference with spacial resolution of 2 $\mu$m.
Coherence length is evaluated by measuring the extension of a
normalized first order correlation function ($g^{(1)}$), i.e.
interference fringe visibility. One of the mirrors of the
interferometer is vibrated mechanically to run interference fringes
on a pin-hole detector.


\begin{figure}

{\centering \resizebox*{0.48\textwidth}{0.55\textheight}
{{\includegraphics{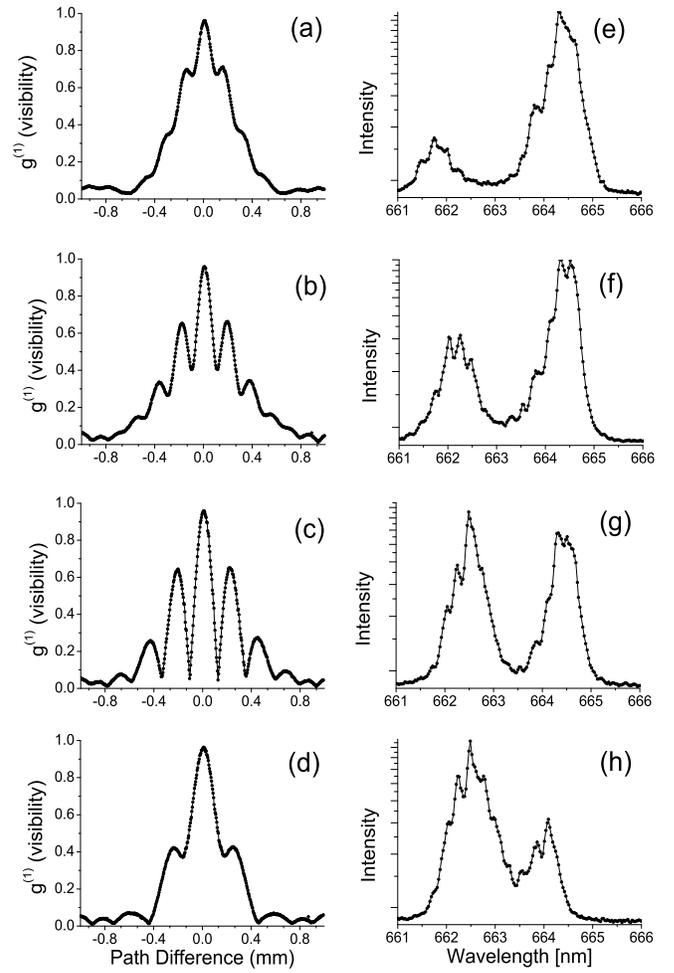}}}
\par}
\caption{\label{vis-experimental} First order correlation function
(a)-(d) and the corresponding optical spectra (e)-(h) are shown for
different feedback polarization angles (a),(e) - 75${^0}$; (b),(f) -
77${^0}$; (c),(g) - 82${^0}$; (d),(h) - 83${^0}$. Notice that the
gap in the optical spectra broadens and moves towards higher
energies for lower polarization angles.}
\end{figure}

\begin{figure}

{\centering \resizebox*{0.45\textwidth}{0.38\textheight}
{{\includegraphics{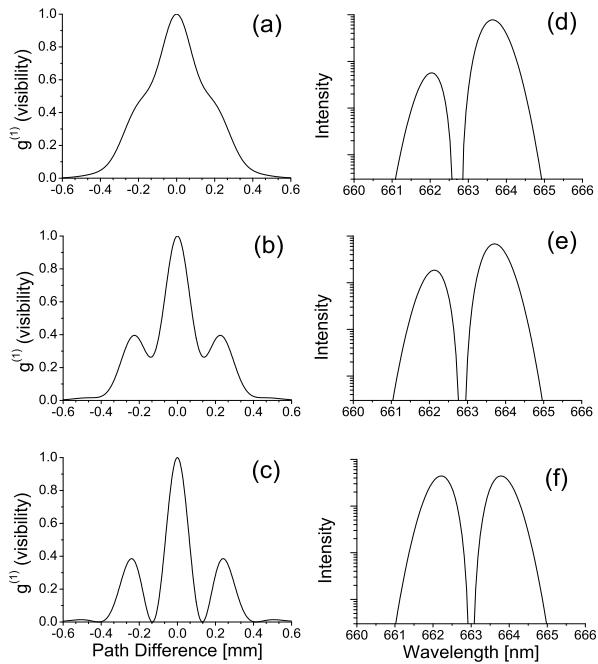}}}
\par}
\caption{\label{vis-theory} Theoretically simulated optical spectra
and their corresponding first order correlation functions.}
\end{figure}

Measurements of $g^{(1)}$ function, for specific values of feedback
polarization rotation angles ($75^{0}$, $77^{0}$, $82^{0}$,
$83^{0}$), are shown in Fig.\ref{vis-experimental}(a)-(d). The
common feature of these measurements is a clearly visible multi-peak
structure with pronounced secondary revivals at distances
significantly shorter than the laser cavity length. For lower and
higher rotation angles the $g^{(1)}$ function indicates no such
sharp multi-peak structure but a smooth and broad single-peak
behavior. In Fig. \ref{vis-experimental}(c) ($82^{0}$) the best
contrast of the multi-peak structure is presented. At zero optical
path difference the fringe visibility is above 95$\%$. Fast initial
fall off, at $\sim$100 $\mu$m, is followed by a revival at $\sim$250
$\mu$m. A number of visible revivals occur at multiples of the first
revival distance. In Fig.\ref{vis-experimental}(a),(b),(d) it is
seen that while the general multi-peak structure of the $g^{(1)}$ is
preserved, each peak broadens and the revival distances decrease
slightly. The observed intriguing multi-peak behavior of the
correlation function suggest a nontrivial gain medium spectrum
profile which is neither Gaussian nor a truncated Gaussian. In fact,
to reproduce these results an additional long wavelength modulation
of the gain medium spectrum is required. In
Fig.\ref{vis-theory}(a)-(c) we show simulations of the $g^{(1)}$
function that bear resemblance to the experimental results
(Fig.\ref{vis-experimental}(a)-(d)). The simulations are based on
initial guess-work of the gain medium spectrum profiles which are
shown in Fig.\ref{vis-theory}(d)-(f). All profiles predict the
existence of a gap in the optical spectrum.

To verify the correctness of our guess-work we measure the optical
spectrum using an optical spectrum analyzer (see Fig.\ref{setup}). A
striking similarity to our predictions is evident in Fig.
\ref{vis-experimental}(e)-(h) where the laser optical spectrum was
measured for different feedback polarization angles. A gap in the
spectrum is clearly visible and, most significantly, its position
and width can be easily altered by a simple rotation of the feedback
polarization angle. At $83^{0}$ (Fig.\ref{vis-experimental}(h)) the
gap is centered at $\sim$663.5 nm and its width is less than 1 nm.
For lower polarization angles the gap widens and moves towards
shorter wavelengths (higher energies). At $75^{0}$
(Fig.\ref{vis-experimental}(e)) it is centered at $\sim$662.9 nm and
its width is $\sim$1.5 nm which is consistent with $\sim$25
consequent forbidden longitudinal modes. For higher and lower
polarization angles the observed spectrum was essentially gapless.
When the gap is located in the middle of the spectrum
(Fig.\ref{vis-experimental}(g)), the best contrast of the multipeak
structure of the first order correlation function is achieved
(Fig.\ref{vis-experimental}(c)).


To interpret our results we assume the condition of slow diffusion
of the gain medium free carriers. Thus a standing wave in the laser
cavity imprints a carrier density grating which, in turn, induces
periodic modulation of the gain medium refractive index. This
modulation imposes a forbidden energy-gap on the laser spectra. In
order to perceive this gap, modulation strength reduction is
necessary (see below). Our experimental approach allows tuning of
the modulation strength in a way similar to the twisted-mode
technique \cite{Evtuhov65}. Two identical counter-propagating beams
create an intensity standing wave when their polarizations are
parallel, whereas a polarization gradient is formed if they are
orthogonal. The latter case produces a standing-wave pattern with no
electric field nodes. If formed in a laser cavity, it erases traces
of spatial hole-burning due to the constant wave intensity, hence
reducing carrier density grating's depth. By rotating the
polarization of the fed-back light, i.e. changing the relative
intensities of the TE-TM polarizations, we effectively regulate the
modulation strength, resulting in a well controlled 1-dimensional
(1D) periodic potential.

To make a quantitative analysis we represent the induced periodic
potential in terms of dielectric function $\varepsilon (x)$:
\begin{equation}
\varepsilon(x)=\varepsilon_{m}(1+\Delta\sin^{2}(2\pi x/a)),
\label{modulation}
\end{equation}
where $\varepsilon_{m}$ is the dielectric constant of the gain
medium, $a=\lambda/2$ is the period of modulation and $\Delta$ is
the modulation amplitude. A 1D single mode Maxwell's equation for
the electric field propagation in the periodically modulated medium
\begin{equation}
\frac{d^{2} E}{dx^{2}}
-\left(\frac{\omega_{0}}{c}\right)^2\varepsilon(x)E=0,
\label{Mathieu-eq}
\end{equation}
where $\omega_{0}=kc$ and $k=2\pi/\lambda$, can be readily reduced
to a familiar form of a well known Mathieu equation:
\begin{equation}
\frac{d^{2} E}{dx'^{2}}
-\frac{1}{4}\varepsilon_{m}((1+\frac{\Delta}{2})-\frac{\Delta}{2}\cos(2x'))E=0,
\label{Mathieu-eq}
\end{equation}
where $x'=2kx$ \cite{omega}.

Let us point out that the carrier density grating in a multimode
laser cavity show more complicated structure than that of a simple
single-mode standing wave. Mode beating produces long-wavelength
spatial variation which periodicity depends on the number of
exciting longitudinal modes \cite{Masoller05}. A single-mode
approach is a good approximation in the case of relatively small
number of modes. From our measurements we estimate that $\sim 25$
modes exist in the cavity, hence producing a modulation periodicity
of $\sim 38$ $\mu$m. This is by an order of magnitude larger than
the optical standing-wave periodicity (0.33 $\mu$m) and therefore
its influence on a band-gap structure is small. However in a case of
extreme multimode condition, mode beating should be considered to
appropriately define the band gap structure.


\begin{figure}

{\centering \resizebox*{0.45\textwidth}{0.27\textheight}
{{\includegraphics{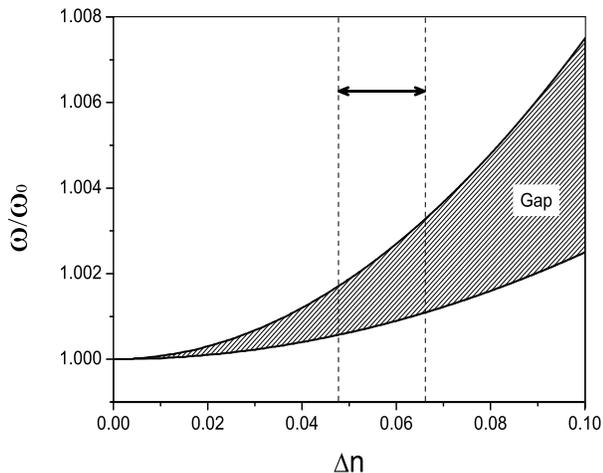}}}
\par}
\caption{\label{band-gap} Band-gap structure for weak modulation
potential (eq.(\ref{modulation})). Bordered by vertical lines is the
region where the gap is experimentally visible in the optical
spectrum (see Fig.\ref{vis-experimental}).}
\end{figure}

We use a text-book solution to the Mathieu equation eigenvalue
problem \cite{Mathieu-spectrum} to build a band-gap structure of the
periodic potential (\ref{modulation}) as a function of the
refractive index modulation depth. In Fig.\ref{band-gap} the
behavior of the first energy gap is shown for weak modulation
depths. The gap quickly widen and depart towards higher frequency as
the modulation becomes stronger. For $\Delta n=0.1$ the lower gap
edge is already $\sim1.5$ nm away from the laser mode frequency and
$\sim3.2$ nm wide which clearly prevents its observation. Only for
weak refractive index modulations it is possible to see the gap and
fill some population in the second band. In Fig.\ref{band-gap} the
range of the gaps observed in the experiment is boarded by vertical
lines that correspond to refractive index modulations between
$\Delta n\approx0.048$ and $\Delta n\approx0.066$. These values are
reasonable and can be obtained in the gain medium. For stronger
potentials (lower polarization angles) the gap continue to widen and
move towards higher energies until the population of the second band
becomes invisible. For smaller potentials (higher polarization
angles) the gap is no more visible and the optical spectrum becomes
narrower.

In fact the observed perturbation of the refractive index is very
small and we can consider a simple perturbation theory to calculate
the energy gap opening at the boundary of the first Brillouin zone
\cite{A&M}. The perturbation theory analysis predicts a gap width
$\Delta \omega/\omega_{0} \approx \Delta/2$. For example, for
$\Delta n\approx0.048$, $\Delta \omega/\omega_{0} \approx (\Delta
n)^2/2 \approx 1.2\times10^{-3}$ which is in excellent agreement
with the result shown in Fig.\ref{band-gap}.

In conclusion we show a regime of distinct multi-mode dynamics in a
semiconductor laser subject to intense variably rotated optical
feedback. Our coherence length and optical spectra measurements
reveal a giant gap in the optical spectrum in a specific range of
the feedback polarization rotation angles. The observed gap is more
than a terahertz wide leading to essentially two-color emission
conditions. We suggest a theoretical approach based on slow carrier
diffusion condition which predicts formation of a carrier density
grating in the cavity. The grating modulates the refractive index of
the gain medium which opens an energy forbidden gap in the optical
spectrum of the laser. We show control on the gap position and width
by means of rotation of the feedback polarization angle indicating
the ability to tune the induced potential.

We note that a similar nonuniform optical spectrum was recently
identified in a short external cavity high power SL under resonant
condition \cite{Peil06}. Finally, the wide optical spectrum,
demonstrated in the experiments, makes the high power SL subjected
to polarization rotated feedback an excellent candidate for using in
applications where a bright incoherent source is required such as
optical tomography and inspection.

This work was supported, in a part, by the Israel Science Foundation
through grant No. 1125/04.


\begin{thebibliography}{99}
\bibitem{Hohl99} A Hohl and A. Gavrielides, Phys. Rev. Lett. \textbf{82}, 1148 (1999).

\bibitem{Heil01} T. Heil, I. Fischer, W. Els\"{a}{\ss}er, and A. Gavrielides,
Phys. Rev. Lett. \textbf{87}, 243901 (2001).

\bibitem{Ushakov04} O. Ushakov, S. Bauer, O. Brox, H.-J.
W\"{u}nsche, and F. Henneberger, Phys. Rev. Lett. \textbf{92},
043902 (2004).

\bibitem{Schikora06} S. Schikora, P. H\"{o}vel, H.-J. W\"{u}nsche,
E. Sch\"{o}ll and F. Henneberger, Phys. Rev. Lett. \textbf{97},
213902 (2006).

\bibitem{Farias05} B. Farias, T. Passerat de Silans, M. Chevrollier,
and M. Ori\'{a}, Phys. Rev. Lett. \textbf{94}, 173902 (2005).

\bibitem{Vemuri04} A.P.A. Fischer, M. Yousefi, D. Lenstra, M.W. Carter
and G. Vemuri, Phys. Rev. Lett. \textbf{92}, 023901 (2004).

\bibitem{LangKobayshi} R. Lang and K. Kobayshi, IEEE J. Quantum
Electron. \textbf{16}, 347 (1980).

\bibitem{Uchida01} A. Uchida, Y. Liu, I. Fischer, P. Davis, and T.
Aida, Phys. Rev. A \textbf{64}, 023801 (2001).

\bibitem{Peil06} M. Peil, I. Fischer, and W. Els\"{a}{\ss}er, Phys.
Rev. A \textbf{73}, 023805 (2006).

\bibitem{Viktorov00} E.A. Viktorov and P. Mandel, Phys. Rev. Lett.
\textbf{85}, 3157 (2000).

\bibitem{Masoller05} C. Masoller, M.S. Torre, and P. Mandel, Phys.
Rev. A \textbf{71}, 013818 (2005).

\bibitem{Serrat03} C. Serrat, S. Prins, and R. Vilaseca. Phys. Rev.
A \textbf{68}, 053804 (2003).

\bibitem{Li98} H. Li, A. Hohl, A. Gavrielides, H. Hou, and K.D.
Choquette, Appl. Phys. Lett. \textbf{72}, 2355 (1998).

\bibitem{Carr01} T.W. Carr, D. Pieroux, and P. Mandel, Phys. Rev. A
\textbf{63}, 033817 (2001).

\bibitem{Evtuhov65} V. Evtuhov, and A. Siegman, Appl. Opt.
\textbf{4}, 142 (1965).

\bibitem{omega} In our simplified model we assume that $\omega_{0}$ is independent
of the refractive index which is generally incorrect. The laser
frequency decreases with the increase of the refractive index. This
dependence should be taken into account in order to build a correct
band-gap structure. Although for a wide optical spectrum of $~3$ nm
(see Fig.\ref{vis-experimental}(e)-(h)) and for $\Delta<<1$ we
expect this correction to be small.


\bibitem{Mathieu-spectrum} M. Abramowitz and I.A. Stegun, \textit{Handbook
of Mathematical functions} (Dover, New York, 1965).

\bibitem{A&M} N.W. Ashcroft and N.D. Mermin, \textit{Solid State Physics}
(Holt, Rinehart $\&$ Winston, New York, 1976).


\end{thebibliography}
\end{document}